\documentclass[aps,nofootinbib,preprint]{revtex4}

\usepackage{graphicx}
\usepackage{graphics}
\usepackage{amsmath}
\usepackage{subfigure}
\usepackage{color}
\usepackage{latexsym}
\usepackage{setspace}
\input{colordvi.tex}

\newcommand{\gtsim}{\protect\raisebox{-0.5ex}{$\:\stackrel{\textstyle >}{\sim}\:$}}

\begin{document}

\begin{flushright}
RESCEU-41/13 \\
YITP-13-39
\end{flushright}

\title{Langevin description of gauged scalar fields in a thermal bath}

\author{Yuhei Miyamoto$^{1,2}$}
\email[Email: ]{miyamoto"at"resceu.s.u-tokyo.ac.jp}

\author{Hayato Motohashi$^{2,3}$}
\email[Email: ]{motohashi"at"resceu.s.u-tokyo.ac.jp}

\author{Teruaki Suyama$^{2}$}
\email[Email: ]{suyama"at"resceu.s.u-tokyo.ac.jp}

\author{Jun'ichi~Yokoyama$^{2,4}$}
\email[Email: ]{yokoyama"at"resceu.s.u-tokyo.ac.jp}

\affiliation{
$^{1}$ Department of Physics, Graduate School of Science, \\
The University of Tokyo, Tokyo 113-0033, Japan \\
$^{2}$ Research Center for the Early Universe (RESCEU), \\
Graduate School of Science, The University of Tokyo, Tokyo 113-0033, Japan \\
$^{3}$ Yukawa Institute for Theoretical Physics, \\
Kyoto University, Kyoto 606-8502, Japan \\
$^{4}$ Kavli Institute for the Physics and Mathematics of the Universe (WPI), \\
The University of Tokyo, Kashiwa, Chiba, 277-8568, Japan}

\pacs{ }

\begin{abstract}
We study the dynamics of the oscillating gauged scalar field
in a thermal bath.
A Langevin type equation of motion of the scalar field,
which contains both dissipation and fluctuation terms,
is derived by using the real-time finite temperature effective action approach.
The existence of the quantum fluctuation-dissipation relation between 
the non-local dissipation term and the Gaussian stochastic noise terms is verified.
We find the noise variables are anti-correlated at equal-time.
The dissipation rate for the each mode is also studied,
which turns out to depend on the wavenumber.

\end{abstract}

\maketitle

\section{Introduction}

Recent advancements in observational technology enable us to 
trace back the history of the Universe.
In particular, observations of the cosmic microwave background, including the latest
results of the Planck mission \cite{Ade:2013zuv}, provide us
with the picture of the Universe at the recombination, 
the subsequent evolution, and a piece of information on the early Universe.
The Universe in a much earlier period, however, is still veiled 
and many models which are built to
explain the physics beyond the energy scale realized in laboratories 
remain unverified.
To select the theory describing our world,
we need not only observational developments but also
more precise theoretical predictions
using fundamental theories of physics.

One of the most interesting phenomena in the early Universe
is the phase transition.
It has provided mechanisms of inflation \cite{R2, old1, old2, new1, new2, Linde:1983gd}, 
called ``old inflation'' \cite{old1, old2} and ``new inflation'' \cite{new1, new2}.
In both of the models inflation is driven by the vaccum energy before the end of the phase transition.
The thermal inflation \cite{Lyth:1995ka}, also caused by the potential energy of the flaton field,
is a relatively short accelerating period after the primordial inflation.
Since it changes the expansion history of the Universe,
not only the moduli and gravitinos are diluted 
but also the primordial gravitational waves are damped \cite{Easther:2008sx}.
On the other hand, collisions of bubbles generated during a phase transition can 
produce gravitational waves \cite{Kamionkowski:1993fg}.
Furthermore, depending on the kinds of the broken symmetry,
various topological defects are expected to be produced.
Among them, line-like topological defects known as cosmic strings can produce gravitational waves \cite{CSGW}
which may be detectable by future experiments \cite{Kuroyanagi:2012wm}.
These examples indicate that the phase transition is a key to
understand high-energy physics and the early Universe.

Precise description of the dynamics of phase transitions is necessary to
compare predictions of each theoretical model with observations.
In many models of the early Universe, phase transitions are controlled by
the expectation value of the scalar fields.
While the effective potential is a useful quantity to derive properties of
the phase transitions that happen quasi-statically,
its use often comes short because of dynamical nature of the scalar fields.
In such cases, we need to directly solve the evolution equations of the
scalar fields derived from the effective action.
It has been shown that 
the behavior of a scalar field in thermal bath can be described by the Langevin equation \cite{Morikawa:1986rp},
which includes stochastic noise terms coming from interactions with other fields in thermal bath.
These noise terms may change the types of phase transitions.
For example, a previous study \cite{Yamaguchi:1996dp} indicates that
the fermionic noise may lead to the phase mixing, which cannot be described by the effective potential.

So far, the effective action and the resultant equation of motion of a scalar field has been studied 
in models where it has self-interaction and interactions with other fermions and scalar bosons 
\cite{Yamaguchi:1996dp, Yokoyama:2004pf}.
Now, it is an interesting project to extend the previous studies to include interactions with gauge fields.
We extend previous analyses to include interactions with gauge fields
using the simplest Abelian gauge theory known as scalar quantum electrodynamics.
Though the hot scalar QED theory has been studied by Ref. \cite{Wang:2000via, Boyanovsky:1998pg}
to study the dynamics of gauge fields,
we focus on the scalar field as a system of interest and treat gauge fields as a hot environment.

The organization of this paper is as follows.
We briefly review the effective action method and apply it 
to the scalar quantum electrodynamics in Section II. 
Actually, this effective action contains the imaginary part.
In Section III, we interpret it as stochastic noises and derive 
a generalized Langevin equation.
We consider the meaning of the equation, 
and explain the validity of this interpretation.
We also show the stochastic property of the noise,
and compare it with the fermionic and scalar bosonic noises 
which have been studied in previous studies. 
The dissipation rate of the each mode is also studied.
We summarize our study and discuss its applicability in Section IV.

\section{Effective action}
As we have mentioned in the inrtoduction, one of our goals is the precise description of phase transitions in gauge theory
which requires our knowledge of effective action for scalar field.
In this paper, we focus on the derivation of the effective action and investigation of
basic properties of the obtained equation of motion for the massive charged scalar field due to 
interactions with gauge fields.
In order to realize phase transition, we need to add self-interaction of the scalar field to
have Higgs mechanism.
We defer the inclusion of the self-interaction to another study.
The simplest approach to describe phase transitions in the hot early Universe is
to analyze a finite-temperature effective potential.
By using the effective potential, we can explain the symmetry restoration at high temperature 
or in the early Universe and the subsequent spontaneous symmetry breaking.
However, since it is derived on the assumption of static, homogeneous field configuration,
it cannot describe the dynamics of phase transitions accurately.
In this section, in order to obtain the equation of motion governing the dynamical phenomenon, 
we calculate the effective action.
Studies so far show that effective action generally contains the imaginary part,
which can be interpreted as the origin of dissipative properties.

\subsection{Settings}
To clarify the role of gauge fields,
we consider the scalar quantum electrodynamics, which is the simplest gauge theory.
Its Lagrangian density is given by
\begin{align}
\mathcal{L}=& D_{\mu}\Phi^{\dagger}D^{\mu}\Phi -m^2\Phi^{\dagger}\Phi -\frac{1}{4}F_{\mu\nu}F^{\mu\nu}  \notag\\
=&\partial_{\mu}\Phi^{\dagger}\partial^{\mu}\Phi -m^2\Phi^{\dagger}\Phi-\frac{1}{4}F_{\mu\nu}F^{\mu\nu} \notag\\
&+ieA_{\mu}(\Phi^{\dagger}\partial^{\mu}\Phi-\Phi\partial^{\mu}\Phi^{\dagger})+e^2A_{\mu}A^{\mu}\Phi^{\dagger}\Phi \,.
\end{align}

After imposing the Coulomb gauge condition $\vec{\nabla} \cdot \vec{A}=0$, 
one can see the Lagrangian density becomes
\begin{align}
\mathcal{L}=&\partial_{\mu}\Phi^{\dagger}\partial^{\mu}\Phi -m^2\Phi^{\dagger}\Phi +\frac{1}{2}\partial_{\mu}\vec{A}_T\cdot\partial^{\mu}\vec{A}_T \notag\\
&-ie\vec{A}_T(\Phi\vec{\nabla}\Phi^{\dagger}-\Phi^{\dagger} \vec{\nabla} \Phi) 
 -e^2\vec{A}_T \cdot \vec{A}_T \Phi^{\dagger}\Phi        \notag\\
 &+\frac{1}{2}(\vec{\nabla}A_0)^2-ieA_0(\Phi \dot\Phi^{\dagger}-\Phi^{\dagger} \dot\Phi)
 +e^2A_0^2\Phi^{\dagger}\Phi
\,.
\end{align}
Here, $\vec{A}_{\rm T}$ means the transverse components, which satisfy $\vec{\nabla} \cdot \vec{A}_{\rm T}=0$.

In the so-called real-time thermal field theory,
we can calculate thermal average using path integral.\footnote{One of the introductory textbooks is \cite{Bellac_text}.}
We can choose the time path so that it consists of three paths:
$(i)$ A path from $t_i\,\,(<0)$ to $-t_i$ on real axis of $t$ (plus contour),
$(ii)$ a path from $-t_i$ to $t_i$ on real axis of $t$ (minus contour),
and $(iii)$ a path from $t_i$ to $t_i-i\beta$, where
$\beta=1/T$ is the inverse of the temperature.
We can neglect the contribution from the third contour when taking $T\rightarrow \infty$ \cite{Landsman:1986uw}.
We denote field variables on the contour ($i$) and ($ii$) by superscripts $+$ and $-$, respectively.
Following Boyanovsky et al. \cite{Boyanovsky:1998pg}, 
thermal propagators for scalar field $\Phi$ and gauge field $\vec{A}_{\rm T}$ are given as follows.

Propagators for scalar field:
\begin{equation}
\langle \Phi^{(a)\dagger}(\vec{x},t)\Phi^{(b)}(\vec{x'},t')  \rangle
=-i\int \frac{d^3k}{(2\pi)^3}G^{ab}_k(t,t')e^{-i\vec{k}\cdot(\vec{x}-\vec{x'})}
\end{equation}
\begin{align}
G^{++}_k(t,t')&=G^>_k(t,t')\Theta(t-t')+G^<_k(t,t')\Theta(t'-t)\\
G^{--}_k(t,t')&=G^>_k(t,t')\Theta(t'-t)+G^<_k(t,t')\Theta(t-t')\\
G^{+-}_k(t,t')&=-G^<_k(t,t')\\
G^{-+}_k(t,t')&=-G^>_k(t,t')
\end{align}
\begin{align}
\label{Ggreater}
G^>_k(t,t')&=\frac{i}{2\omega_k}\Big[(1+n_k)e^{-i\omega_k (t-t')} +n_ke^{+i\omega_k (t-t')} \Big]\\
\label{Gless}
G^<_k(t,t')&=\frac{i}{2\omega_k}\Big[n_ke^{-i\omega_k (t-t')} +(1+n_k)e^{+i\omega_k (t-t')} \Big]
\end{align}
\begin{equation}
\omega_k=\sqrt{|\vec{k}|^2+m^2},\,n_k=\frac{1}{e^{\beta \omega_k}-1}
\end{equation}

Propagators for gauge field:
\begin{equation}
\langle A_{Ti}^{(a)}(\vec{x},t)A_{Tj}^{(b)}(\vec{x'},t')  \rangle
=-i\int \frac{d^3k}{(2\pi)^3}{\mathcal G}^{ab}_{k\,ij}(t,t')e^{-i\vec{k}\cdot(\vec{x}-\vec{x'})}
\end{equation}
\begin{align}
{\mathcal G}^{++}_{k\,\,ij}(t,t')&={\mathcal P}_{ij}(\vec{k})\Big[g^>_k(t,t')\Theta(t-t')+g^<_k(t,t')\Theta(t'-t) \Big]\\
{\mathcal G}^{--}_{k\,\,ij}(t,t')&={\mathcal P}_{ij}(\vec{k})\Big[g^>_k(t,t')\Theta(t'-t)+g^<_k(t,t')\Theta(t-t') \Big]\\
{\mathcal G}^{+-}_{k\,\,ij}(t,t')&=-{\mathcal P}_{ij}(\vec{k})g^<_k(t,t')\\
{\mathcal G}^{-+}_{k\,\,ij}(t,t')&=-{\mathcal P}_{ij}(\vec{k})g^>_k(t,t')
\end{align}
\begin{align}
\label{ggreater}
g^>_k(t,t')&=\frac{i}{2k}\Big[(1+N_k)e^{-ik (t-t')} +N_ke^{+ik (t-t')} \Big]\\
\label{gless}
g^<_k(t,t')&=\frac{i}{2k}\Big[N_ke^{-ik (t-t')} +(1+N_k)e^{+ik (t-t')} \Big]
\end{align}
\begin{equation}
k=\sqrt{|\vec{k}|^2},\quad N_k=\frac{1}{e^{\beta k}-1},\quad
{\mathcal P}_{ij}(\vec{k})=\delta_{ij}-\frac{k_ik_j}{k^2}
\end{equation}

The generating functional of Green's function is
\begin{equation}
\int \mathcal{D}\vec{A}_T\mathcal{D}\Phi \mathcal{D}\Phi^{\dagger} \exp
\left[ i\int d^4x\,\left\{ \mathcal{L}[\vec{A}^{(+)}_T,\, \Phi^{(+)},\, \Phi^{\dagger(+)}]-
\mathcal{L}[\vec{A}^{(-)}_T,\, \Phi^{(-)},\, \Phi^{\dagger(-)}] \right\}\right] \,.
\end{equation}

\subsection{Perturbative expansion}
\begin{figure}[htbp]
 \begin{center}
  \includegraphics[width=130mm]{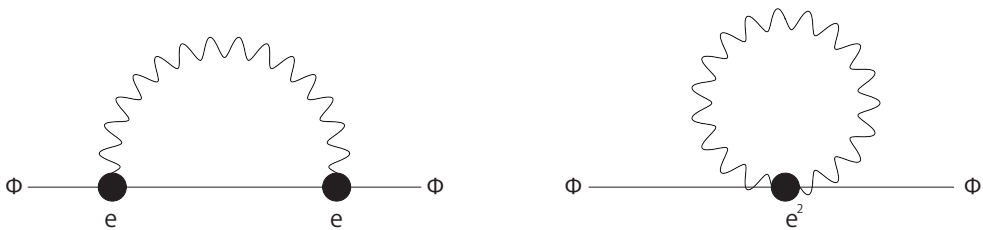}
 \end{center}
  \begin{minipage}[c]{140mm}
 \begin{spacing}{1}
 \caption{
 \small $\mathcal{O}(e^2)$ 1PI diagrams. The solid/wavy line represents the scalar/photon propagator, respectively.
   The left diagram produces non-local terms in the effective action.
 The right diagram gives a thermal correction to the mass term, which is proportional to $T^2$.\label{diagram}}
 \end{spacing}
 \end{minipage}
\end{figure}

 Calculating the effective action corresponds to the summation of one-particle-irreducible diagrams.
 Practically, effective action can be obtained only by means of perturbative expansion 
 in terms of the gauge coupling constant $e$, which we adopt in our study.
 The lowest non-trivial contributions to the effective action appear at the second order of coupling constant $e$.
 At this order, there are two relevant diagrams, which are shown in FIG. \ref{diagram}.
In addition to these 1PI diagrams, we have to rewrite $A_0$ using its Euler-Lagrange equation as
\begin{align}
A_0(x)&=-\frac{1}{\Delta}\rho(x) +{\cal O}(e^2)\\
\rho&\equiv ie(\Phi \dot\Phi^{\dagger}-\Phi^{\dagger} \dot\Phi)
\end{align}
and take the interaction ${\cal L}_{\rm int}=\frac{1}{2}\rho \frac{1}{\Delta}\rho$ into account.

The contribution of the left diagram in FIG. \ref{diagram} to the effective action $\Gamma$ is
\begin{align}
\Gamma \supset
&+4ie^2\int d^4x_1d^4x_2 \,\Big< A^{(+)}_{Ti}(x_1)A^{(+)}_{Tj}(x_2) \Big>
\Big< \partial_i \Phi^{\dagger(+)}(x_1) \partial_j\Phi^{(+)}(x_2)  \Big> \Phi^{\dagger(+)}(x_2)\Phi^{(+)}(x_1)\notag\\
&+4ie^2\int d^4x_1d^4x_2 \,\Big< A^{(-)}_{Ti}(x_1)A^{(-)}_{Tj}(x_2) \Big>
\Big< \partial_i \Phi^{\dagger(-)}(x_1) \partial_j\Phi^{(-)}(x_2)  \Big> \Phi^{\dagger(-)}(x_2)\Phi^{(-)}(x_1)\notag\\
&-4ie^2\int d^4x_1d^4x_2 \,\Big< A^{(-)}_{Ti}(x_1)A^{(+)}_{Tj}(x_2) \Big>
\Big< \partial_i \Phi^{\dagger(-)}(x_1) \partial_j\Phi^{(+)}(x_2)  \Big> \Phi^{\dagger(+)}(x_2)\Phi^{(-)}(x_1)\notag\\
&-4ie^2\int d^4x_1d^4x_2 \,\Big< A^{(+)}_{Ti}(x_1)A^{(-)}_{Tj}(x_2) \Big>
\Big< \partial_i \Phi^{\dagger(+)}(x_1) \partial_j\Phi^{(-)}(x_2)  \Big> \Phi^{\dagger(-)}(x_2)\Phi^{(+)}(x_1)\,,
\end{align}
and the right diagram contributes
\begin{equation}
\Gamma \supset -e^2\int d^4x \Big[\Big< A^{(+)}_{Ti}(x)A^{(+)}_{Ti}(x) \Big>\Phi^{\dagger(+)}(x)\Phi^{(+)}(x)
-\Big< A^{(-)}_{Ti}(x)A^{(-)}_{Ti}(x) \Big>\Phi^{\dagger(+)}(x)\Phi^{(-)}(x)\Big]\,.
\end{equation}
This local term gives a thermal correction to the mass term.
\footnote{Here we omit the divergent part, which is to be cancelled by a mass counter term since it exists even at zero temperature.}
\begin{equation}
\Big< A^{(+)}_{Ti}(x)A^{(+)}_{Ti}(x) \Big>=\Big< A^{(-)}_{Ti}(x)A^{(-)}_{Ti}(x) \Big>=\frac{T^2}{6}
\end{equation}

From the interaction originally expressed by $A_0$, we obtain 
\begin{align}
\Gamma \supset
&-e^2\int\frac{d^4k}{(2\pi)^4}\frac{d^4p}{(2\pi)^4}\frac{d^4q}{(2\pi)^4} \frac{1}{|\vec{p}|^2}(2k_0-p_0)(2q_0-p_0)  \notag\\
&\times \Big[ \Big< \tilde{\Phi}^{\dagger(+)}(k-p) \tilde{\Phi}^{(+)}(q)\Big>  \tilde{\Phi}^{(+)}(k)\tilde{\Phi}^{\dagger(+)}(q+p) \notag\\
&\quad -\Big< \tilde{\Phi}^{\dagger(-)}(k-p) \tilde{\Phi}^{(-)}(q)\Big>  \tilde{\Phi}^{(-)}(k) \tilde{\Phi}^{\dagger(-)}(q+p) \Big]\,.
\end{align}

It is convenient to replace $\Phi^{(\pm)}$ with new variables
\begin{equation}
\Phi^{(\pm)}= \phi_{c} \pm \frac{1}{2}\phi_{\Delta} \,.
\end{equation}

Finally the effective action $\Gamma$ incorporating these two diagrams 
and $A_0$ terms
up to the second order in $e$ becomes
\begin{align}
\Gamma=&\int d^4x\, \Big[ \phi_{\Delta}^{\dagger}(x)\left(-\partial_{\mu}\partial^{\mu}-m^2-e^2\frac{T^2}{6}\right)\phi_{c}(x) \notag \\
&\qquad\quad +\phi_{\Delta}(x)\left(-\partial_{\mu}\partial^{\mu}-m^2-e^2\frac{T^2}{6}\right)\phi_{c}(x)^{\dagger} \Big] \notag \\
&-4ie^2\int d^4x_1d^4x_2 \int \frac{d^3p_1}{(2\pi)^3}\frac{d^3p_2}{(2\pi)^3}e^{-i(\vec{p_1}+\vec{p_2})\cdot(\vec{x_1}-\vec{x_2})}
\mathcal{P}_{ij}(\vec{p_1})p_{2i}p_{2j}  \Theta(t_2-t_1) \notag \\
&\qquad \big[g^<_{p_1}(t_1,t_2)G^<_{p_2}(t_1,t_2)-g^>_{p_1}(t_1,t_2)G^>_{p_2}(t_1,t_2)  \big]
\left( \phi_{c}^{\dagger}(x_1)\phi_{\Delta}(x_2) + \phi_{c}(x_1)\phi_{\Delta}^{\dagger}(x_2)  \right)\notag\\
&-2ie^2\int d^4x_1d^4x_2 \int \frac{d^3p_1}{(2\pi)^3}\frac{d^3p_2}{(2\pi)^3}e^{-i(\vec{p_1}+\vec{p_2})\cdot(\vec{x_1}-\vec{x_2})}
\mathcal{P}_{ij}(\vec{p_1})p_{2i}p_{2j}   \notag \\
&\qquad \big[g^<_{p_1}(t_1,t_2)G^<_{p_2}(t_1,t_2)+g^>_{p_1}(t_1,t_2)G^>_{p_2}(t_1,t_2)  \big]
\phi_{\Delta}^{\dagger}(x_1)\phi_{\Delta}(x_2) \notag\\
&+\Gamma_{A_0}\,,
\end{align}
\begin{align}
\Gamma_{A_0}=
&ie^2\int\frac{d^4k}{(2\pi)^4}\frac{d^4q}{(2\pi)^4}\frac{(k_0+q_0)^2}{|\vec{k}-\vec{q}|^2}\int d\tau e^{iq_0\tau}(G^>_q(\tau)+G^<_q(\tau)) \notag\\
&\qquad \times \frac{1}{2}\left( \tilde{\phi_c}(k)\tilde{\phi_\Delta}^{\dagger}(k)+\tilde{\phi_\Delta}(k)\tilde{\phi_c}^{\dagger}(k) \right)  \notag\\
\equiv&\int\frac{d^4k}{(2\pi)^4}\left( \tilde{\phi_c}(k)\tilde{\phi_\Delta}^{\dagger}(k)+\tilde{\phi_\Delta}(k)\tilde{\phi_c}^{\dagger}(k) \right)\tilde{f}_{A_0}(k)
\end{align}

Let us show that the imaginary part of the non-local terms 
which come from diagrams in FIG. \ref{diagram} is nonzero. 
First, both of the integrands are invariant under replacements 
$\vec{p_1} \rightarrow -\vec{p_1}$ and 
$\vec{p_2} \rightarrow -\vec{p_2}$ respectively.
This property allows us to replace 
$e^{-i(\vec{p_1}+\vec{p_2})\cdot(\vec{x_1}-\vec{x_2})}$
with $\cos \left[(\vec{p_1}+\vec{p_2})\cdot(\vec{x_1}-\vec{x_2})\right]$, which is real.
Second, from Eqs. (\ref{Ggreater}), (\ref{Gless}), (\ref{ggreater}), and (\ref{gless}),
we note that $g^<G^<-g^>G^>$ is purely imaginary and $g^<G^<+g^>G^>$ is real.
Thus, the first non-local term is real and the second one is purely imaginary.
In a similar way, we can see that $\Gamma_{A_0}$ is a real functional.
We will explain how to interpret the imaginary part of the effective action in the next section.

\section{Langevin equation and Noise properties}
 In the previous section, we have seen the effective action for the scalar field contains an imaginary part 
 as with the case of pure scalar theory in thermal environment. 
 It can be written as 
 \begin{equation}
 i\Gamma \supset -\int d^4x_1d^4x_2\,{\cal N}(x_1-x_2)\left(\phi_{\Delta \,{\rm R}}(x_1)\phi_{\Delta\,{\rm R}}(x_2)+\phi_{\Delta\,{\rm I}}(x_1)\phi_{\Delta\,{\rm I}}(x_2)\right)\,,
 \end{equation}
 where
 \begin{align}
\label{Nrealspace}
 {\cal N}(x_1-x_2)=&-2e^2\int \frac{d^3p_1}{(2\pi)^3}\frac{d^3p_2}{(2\pi)^3}e^{-i(\vec{p_1}+\vec{p_2})\cdot(\vec{x_1}-\vec{x_2})}
\mathcal{P}_{ij}(\vec{p_1})p_{2i}p_{2j}  \notag\\
&\qquad \big[g^<_{p_1}(t_1,t_2)G^<_{p_2}(t_1,t_2)+g^>_{p_1}(t_1,t_2)G^>_{p_2}(t_1,t_2)  \big]\,.
 \end{align}
$\phi_{\Delta\,{\rm R/I}}$ are the real/imaginary part of $\phi_{\Delta}$, respectively.
 Now we are going to rewrite and interpret it as stochastic noise terms.

\subsection{Mathematical transformation}
As in previous studies\cite{Morikawa:1986rp,Yamaguchi:1996dp,Yokoyama:2004pf}, we rewrite the imaginary part 
by using Gaussian integral formula
\begin{align}
\label{gaussianformula}
&\exp\left[ - \int d^4xd^4y \, \varphi(x)M(x,y) \varphi(y)\right] \notag\\
\propto&\int \mathcal{D}\xi \exp \left[-\frac{1}{4}\int d^4x d^4y \, \xi(x)M^{-1}(x,y)\xi(y)+i\int d^4x\,\xi(x)\varphi(x)\right] \,,
\end{align}
and interpret the integration over $\xi$ as an ensemble average,
where $\xi$ is regarded as a stochastic Gaussian variable.

This formula is not applicable to arbitrary $M(x,y)$.
Just like one dimensional Gaussian integral $\int dx\, \exp [-ax^2]$ requires $a>0$,
all of the eigenvalues of $M$ should be positive.
After rewriting with Fourier transformations
\begin{equation}
\int d^4x d^4y \, \xi(x)M^{-1}(x,y)\xi(y)=\int \frac{d^4 k}{(2\pi)^4} \tilde{M}^{-1}(k) |\tilde{\xi}(k)|^2  \,,
\end{equation}
we notice that $\tilde{M}(k)$ should be positive.
The Fourier transformation of $\cal N$
with respect to $t-t'$ and $\vec{x}-\vec{x}'$ is
\begin{align}
\label{noise kernel in kspace}
\tilde{\cal N}(\omega, \vec{k})=& 2e^2 \int \frac{d^3p_1}{(2\pi)^3}\frac{d^3p_2}{(2\pi)^3}\mathcal{P}_{ij}(\vec{p_1})p_{2i}p_{2j}(2\pi)^3
\delta(\vec{k}-\vec{p_1}-\vec{p_2})   \notag\\
&\quad\frac{2\pi}{2p_12\omega_{p_2}}\Big[ \bigl\{(1+N_{p_1})(1+n_{p_2})+N_{p_1}n_{p_2}\bigr\}\delta(\omega-p_1-\omega_{p_2})\notag\\
&\qquad\quad\quad +\bigl\{(1+N_{p_1})n_{p_2}+N_{p_1}(1+n_{p_2})\bigr\}\delta(\omega-p_1+\omega_{p_2}) \notag\\
&\qquad\quad\quad +\bigl\{N_{p_1}(1+n_{p_2})+(1+N_{p_1})n_{p_2}\bigr\}\delta(\omega+p_1-\omega_{p_2}) \notag\\
&\qquad\quad\quad +\bigl\{N_{p_1}n_{p_2}+(1+N_{p_1})(1+n_{p_2})\bigr\}\delta(\omega+p_1+\omega_{p_2})\Big]\,. 
\end{align}
Clearly, this is positive for any $(\omega,\vec{k})$, and thus
this expression ensures us 
that we can use the formula (\ref{gaussianformula}) and rewrite the effective action with stochastic noise terms.
Finally we obtain
\begin{align}
e^{i\Gamma}=& \int {\mathcal D}\xi_a{\mathcal D}\xi_b \,P[\xi_a]P[\xi_b]\,\exp\Big[i\Gamma_{\rm real}+i\int d^4x (\xi_a(x)\phi_{\Delta\,{\rm R}}(x)+\xi_b(x)\phi_{\Delta\,{\rm I}}(x))\Big] \notag\\
=& \int {\mathcal D}\xi_a{\mathcal D}\xi_b \,P[\xi_a]P[\xi_b]\,\exp\Big[i\Gamma_{\rm real}+i\int d^4x (\xi^{\dagger}(x)\phi_{\Delta}(x)+\xi(x)\phi_{\Delta}^{\dagger}(x))\Big]\notag\\
\equiv& \int {\mathcal D}\xi_a{\mathcal D}\xi_b \,P[\xi_a]P[\xi_b]\,\exp\Big[ iS_{\rm eff} \Big]\,,
\end{align}
where 
\begin{align}
P[\xi_a]=&\exp \Big[ -\frac{1}{4}\int d^4xd^4y\, \xi_a(x){\cal N}^{-1}(x-y)\xi_a(y) \Big]\notag\\
P[\xi_b]=&\exp \Big[ -\frac{1}{4}\int d^4xd^4y\, \xi_b(x){\cal N}^{-1}(x-y)\xi_b(y) \Big]\notag\\
\xi=&\frac{1}{2}\xi_a +i\frac{1}{2}\xi_b= \xi_{\rm R}+i\xi_{\rm I}  \,.
\end{align}
Now we have a real action $S_{\rm eff}$ containing stochastic noise terms.
The two-point correlation function of $\xi$ is given by

\begin{equation}
\langle \xi_{\rm R}(x_1)\xi_{\rm R}(x_2)\rangle
=\langle \xi_{\rm I}(x_1)\xi_{\rm I}(x_2)\rangle
=\frac{1}{2}{\mathcal N}(x_1-x_2)
\end{equation}
\begin{equation}
\langle \xi_{\rm R}(x_1)\xi_{\rm I}(x_2)\rangle=0
\end{equation}
\begin{equation}
\langle \xi(x_1)\xi^{\dagger}(x_2)\rangle={\mathcal N}(x_1-x_2) \,.
\end{equation}

\subsection{Validity of interpretation and the fluctuation-dissipation relation}

As we saw in the previous section, we obtain a real effective action $S_{\rm eff}$
by introducing noise terms. 
This real action leads to the following Langevin type equation of motion
\begin{align}
&\left(\Box +m^2 +e^2\frac{T^2}{6}\right)\phi_c(x)
-\int d^4x'\, f_{A_0}(x-x')\phi_c(x') \notag\\
&+4ie^2\int_{-\infty}^{t} dt'\int d^3x' \int \frac{d^3p_1}{(2\pi)^3}\frac{d^3p_2}{(2\pi)^3}e^{-i(\vec{p_1}+\vec{p_2})\cdot(\vec{x'}-\vec{x})}
\mathcal{P}_{ij}(\vec{p_1})p_{2i}p_{2j} \notag \\
&\qquad \big[g^<_{p_1}(t',t)G^<_{p_2}(t',t)-g^>_{p_1}(t',t)G^>_{p_2}(t',t)  \big] \phi_c(t',\vec{x'})\notag\\
&=\xi(x)
\,.
\end{align}

Now let us consider its validity.
The right hand side, $\xi$, kicks or fluctuates the mean field and 
supplies energy to it from the thermal bath.
On the other hand, the last term on the left hand side represents the friction,
which dissipates the energy of the mean field $\phi_c$ into the bath.
This non-local memory term can be formally written as
\begin{equation}
\int_{-\infty}^{t}dt'\int d^3x \,C(x-x')\phi(x')\,.
\end{equation} 
The equation of motion in the Fourier space is
\begin{align}
&(-\omega^2+k^2+m^2)\tilde{\phi}(\omega,\vec{k})\notag\\
+&\left(e^2\frac{T^2}{6}-\tilde{f}_{A_0}(\omega,\vec{k})+\int\frac{d\omega'}{2\pi}\frac{\cal{P}}{\omega-\omega'}i\tilde{C}(\omega',\vec{k})\right) \tilde{\phi}(\omega,\vec{k})\notag\\
+&\frac{1}{2}\tilde{C}(\omega,\vec{k}) \tilde{\phi}(\omega,\vec{k})=\tilde{\xi}(\omega,\vec{k})\,,
\end{align}
Note that $\tilde{C}$ is purely imaginary and $\tilde{f}_{A_0}$ is real,
so all the coefficients of $\tilde{\phi}$ in the second line are real.
We interpret them as corrections to the free part, the first line.
The terms in the third line can be interpreted as dissipation and fluctuation.

The imaginary part of the Fourier transformation of the memory kernel $C(t, \vec{x})$ is
\begin{align}
\label{memorykernel in kspace}
{\rm Im}\tilde{C}(\omega, \vec {k})= -4e^2 \int \frac{d^3p_1}{(2\pi)^3}\frac{d^3p_2}{(2\pi)^3}\mathcal{P}_{ij}&(\vec{p_1})p_{2i}p_{2j}(2\pi)^3
\delta^{(3)}(\vec{k}-\vec{p_1}-\vec{p_2})   \notag\\
\times \frac{2\pi}{2p_12\omega_{p_2}}\Big[ &\bigl\{(1+N_{p_1})(1+n_{p_2})-N_{p_1}n_{p_2}\bigr\}\delta(\omega-p_1-\omega_{p_2})\notag\\
 +&\bigl\{(1+N_{p_1})n_{p_2}-N_{p_1}(1+n_{p_2})\bigr\}\delta(\omega-p_1+\omega_{p_2}) \notag\\
 +&\bigl\{N_{p_1}(1+n_{p_2})-(1+N_{p_1})n_{p_2}\bigr\}\delta(\omega+p_1-\omega_{p_2}) \notag\\
 +&\bigl\{N_{p_1}n_{p_2}-(1+N_{p_1})(1+n_{p_2})\bigr\}\delta(\omega+p_1+\omega_{p_2})\Big]\,. 
\end{align}
Now we have collected all the ingredients necessary for showing the fluctuation-dissipation relation.
Expecting that the scalar field and the gauge field reach some equilibrium state,
we start our analysis by using finite temperature propagators.
In order for a system to achieve and keep thermal equilibrium, 
there is a necessary condition between noise terms and the memory term, which is the fluctuation-dissipation relation.
Mathematically, it is written as
\begin{equation}
\label{FDR}
 \frac{\tilde{\cal N}(\omega, \vec k)}{\frac{-1}{\omega}{\rm Im}\tilde{C}(\omega, \vec k)}
=\frac{\omega}{2} \frac{e^{\beta \omega}+1}{e^{\beta \omega}-1}
=\omega \left( \frac{1}{2} +n_{\omega} \right) \,.
\end{equation}
It is straightforward to check that this relation indeed holds
 in our case.\footnote{Owing to delta functions, we can factor out the ratio $\frac{e^{\beta \omega}+1}{e^{\beta \omega}-1}$
without performing complicated integrals in (\ref{noise kernel in kspace}) and (\ref{memorykernel in kspace}). }
This is the quantum fluctuation-dissipation relation \cite{Yokoyama:2004pf, Calzetta:1999xh, Calzetta Hu textbook, Berera:2007qm, Greiner:1996dx}.
In light of this fact, we conclude that
the introduction of noise terms is not just a mathematical trick
but a meaningful transformation to bring out physics.

\subsection{Property of the stochastic noise}

We now show the properties of the noise.
From Eq. (\ref{Nrealspace}), we see that
the spatial noise correlation is expressed as
\begin{align}
& \langle \xi(\vec{x_1},t)\xi^{\dagger}(\vec{x_2},t)\rangle \notag\\
=&\frac{e^2}{2} \int{\frac{d^3p_1}{(2\pi)^3}}\int{\frac{d^3p_2}{(2\pi)^3}}
e^{\displaystyle-i(\vec{p_1}+\vec{p_2})\cdot(\vec{x_1}-\vec{x_2})}
\mathcal{P}_{ij}(\vec{p_1})p_{2i}p_{2j}\frac{1}{p_1 \omega_{p_2}}(1+2N_{p_1})(1+2n_{p_2}) \,.
\end{align}
We divide it as
\begin{align}
\label{spatialcorrelationwithalphabetagamma}
\langle \xi(\vec{x_1},t)\xi^{\dagger}(\vec{x_2},t)\rangle 
=&\frac{e^2}{2} (\alpha_{ij}-\beta_{ij})\gamma_{ij}\notag\\
=&\frac{e^2}{2} \Bigg[ -\alpha(r)\left(\gamma''(r)+\frac{2}{r}\gamma'(r)\right) -\frac{2}{r^2}\beta'(r)\gamma'(r)-\beta''(r)\gamma''(r)   \Bigg] \,,
\end{align}
where
\begin{align}
\alpha_{ij}=& \int{\frac{d^3p_1}{(2\pi)^3}}e^{\displaystyle-i\vec{p_1}\cdot\vec{r}}\,  \frac{1}{p_1}\left(1+\frac{2}{e^{\beta p_1}-1}\right) \delta_{ij} \equiv \alpha(r)\delta_{ij} \,,\\
\beta_{ij}=&  \int{\frac{d^3p_1}{(2\pi)^3}}e^{\displaystyle-i\vec{p_1}\cdot\vec{r}}\,  \frac{p_{1i}p_{1j}}{p^3_1}\left(1+\frac{2}{e^{\beta p_1}-1}\right) 
\equiv -\frac{\partial}{\partial r_i}\frac{\partial}{\partial r_j} \beta(r)  \,,\\
\gamma_{ij}=& \int{\frac{d^3p_2}{(2\pi)^3}}e^{\displaystyle-i\vec{p_2}\cdot\vec{r}}\,  \frac{p_{2i}p_{2j}}{\omega_{p_2}}\left(1+\frac{2}{e^{\beta \omega_{p_2}}-1}\right) 
\equiv-\frac{\partial}{\partial r_i}\frac{\partial}{\partial r_j} \gamma(r) \,,
\end{align}
and we use $r=|\vec{r}|=|\vec{x_1}-\vec{x_2}|$.

After some calculations, we obtain the following expressions.
\begin{align}
\label{alpha(r)}
\alpha(r)=&\frac{1}{2\pi r\beta}{\rm coth} \left(\frac{r}{\beta}\pi \right)  \\
\beta'(r)=& -\frac{1}{2\pi^2 r}+\sum_{n=1}^{\infty}\frac{-r+n\beta {\rm Arccot}(\frac{n\beta}{r})}{\pi^2r^2} \\
\beta''(r)=&\frac{1}{\pi^2r^2}-\frac{1}{2\pi r \beta}{\rm coth}(\frac{r}{\beta}\pi)
+\sum_{n=1}^{\infty}\frac{2}{\pi^2r^3}\Big[r-n\beta {\rm Arccot}(\frac{n\beta}{r})\Big]  \\
\gamma'(r)=& -\frac{m^2}{2\pi^2r}K_2(mr)-\frac{m^2r}{\pi^2}\sum_{n=1}^{\infty}\frac{1}{r^2+n^2\beta^2}K_2(m\sqrt{r^2+n^2\beta^2}) \\
\gamma''(r)=&-\frac{m^2}{2\pi^2}\Big[\frac{1}{r^2}K_2(mr)-\frac{m}{r} K_3(mr)\Big]\notag\\
\label{gammadoubleprime}
&-\frac{m^2}{\pi^2}\sum_{n=1}^{\infty}\Big[\frac{1}{r^2+n^2\beta^2}K_2(m\sqrt{r^2+n^2\beta^2})-\frac{mr^2}{(r^2+n^2\beta^2)^{3/2}}K_3(m\sqrt{r^2+n^2\beta^2}) \Big] 
\end{align}
Here $K_\nu(z)$ is the modified Bessel function of $\nu$-th order.

Though the expression of the noise correlation function is quite cumbersome for general case,
and numerical computation is the only feasible way to evaluate it,
it reduces to a fairly concise form in some limiting cases.
First, in the short distance limit, we find 
\begin{equation}
\label{short}
\langle \xi(\vec{x_1},t)\xi^{\dagger}(\vec{x_2},t)\rangle \,\simeq \,
-\frac{3e^2}{2\pi^4r^6}\Theta(r) +\frac{e^2}{4\pi^4r^5}\delta(r)\,.
\end{equation}
For the derivation of this expression, see Appendix A.
Here we define the step function as
\begin{equation}
\Theta(x)=
\left\{
\begin{array}{l}
0\quad x\leq0 \\
1\quad x>0\,.
\end{array}
\right.
\end{equation}

On the other hand, we obtain the following behavior in the long distance limit $r\gg \beta, \frac{1}{m}$.
\begin{equation}
\label{long massive}
\langle \xi(\vec{x_1},t)\xi^{\dagger}(\vec{x_2},t)\rangle \,\simeq \,
-\frac{e^2m^2}{8\pi^2 \beta^2 r^2}e^{-mr}
\end{equation}
If the scalar field is massless, we obtain
\begin{equation}
\label{long massless}
\langle \xi(\vec{x_1},t)\xi^{\dagger}(\vec{x_2},t)\rangle \,\simeq \,
-\frac{e^2}{8\pi^2\beta^2 r^4}\,.
\end{equation}
For the derivation of these expressions, see Appendix B.

\begin{figure}[!h]
 \begin{center}
  \includegraphics[width=140mm]{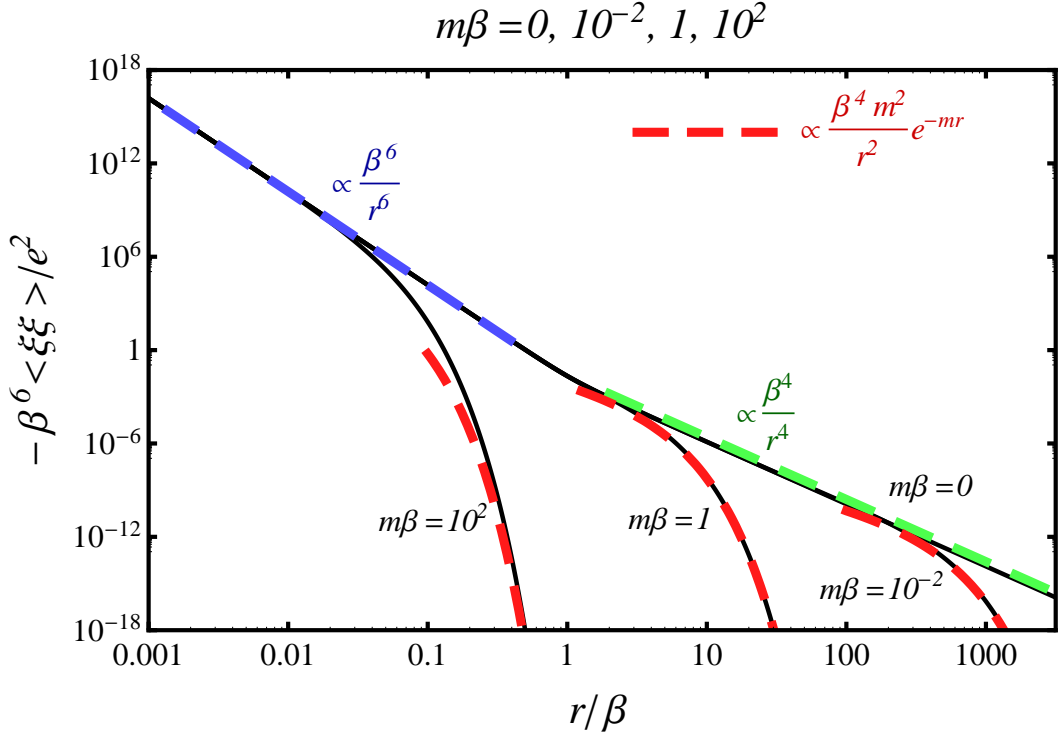}
 \end{center}
   \begin{minipage}[c]{140mm}
 \begin{spacing}{1}
 \caption{Noise spatial correlations for various mass values.
 The solid black line represents the exact expression (\ref{spatialcorrelationwithalphabetagamma})
 with Eqs. (\ref{alpha(r)}) $\sim$ (\ref{gammadoubleprime}).
 The dashed blue, red, and green lines correspond to 
 the analytical approximations (\ref{short}), (\ref{long massive}), and (\ref{long massless}).
  For $r<\frac{1}{m}$, correlations obey power-law decay.
 They start to decay exponentially when $r$ exceeds $\frac{1}{m}$.\label{fig2}}
  \end{spacing}
 \end{minipage}
\end{figure}

We show the spatial correlation for various masses in FIG. \ref{fig2}.
As the approximate expression (\ref{long massive}) shows,
the noise correlation is exponentially 
suppressed at $r\gtsim \frac{1}{m}$ and monotonically approaches zero.
Asymptotically, the noise correlation obtained by numerical evaluation
is consistent with the above simple expressions obtained analytically.
We see that the noise in this model shows anti-correlation,
which is different from the previous study \cite{Yamaguchi:1996dp}.

\subsection{Dissipation rate}
The Langevin equation provides not only fluctuations to the scalar field 
but also its dissipation.

\begin{figure}[]
 \begin{center}
  \includegraphics[width=120mm]{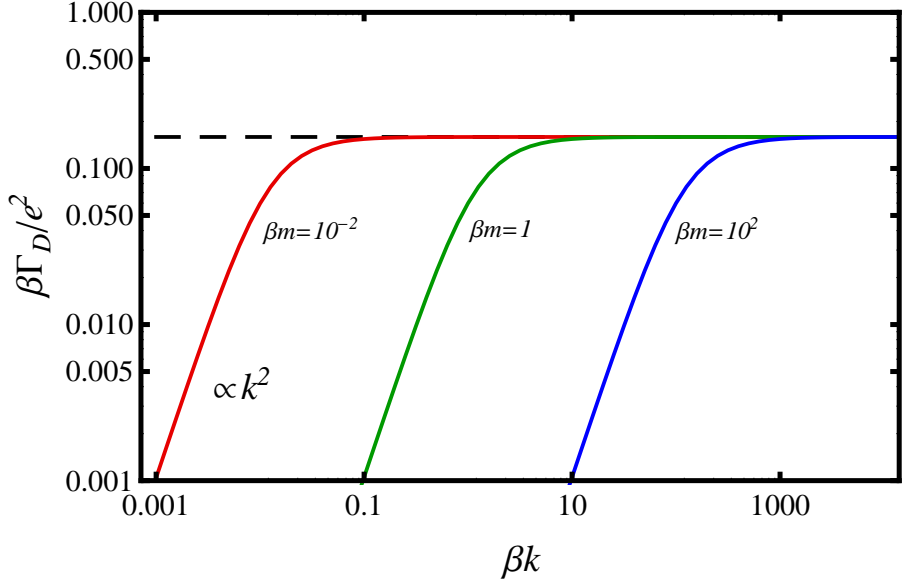}
 \end{center}
   \begin{minipage}[c]{140mm}
 \begin{spacing}{1}
 \caption{The $k$-dependence of the dissipation rate (\ref{dissipationratee2}).
 The dashed black line represents $\beta \Gamma_{\rm D}=e^2/2\pi$.
 The red, green, blue line represents the dissipation rate for 
 $\beta m=10^{-2},\,1,\,10^2$, respectively.
 For $k>m$, the dissipation rate is almost independent of $k$.
 However, for $k<m$, it is proportional to $k^2$.
  \label{dissipationrate}}
  \end{spacing}
 \end{minipage}
\end{figure}

According to Ref. \cite{Yokoyama:2004pf, Greiner:1996dx},
for the scalar field described by the equation
\begin{equation}
(-\omega^2+M_{k,\omega}^2)\tilde{\phi}(\omega,\vec{k})+\frac{1}{2}\tilde{C}(\omega,\vec{k})\tilde{\phi}(\omega,\vec{k})
=\tilde{\xi}(\omega,\vec{k})\,,
\end{equation}
the dissipation rate of the $k$-mode oscillation is given by 
\begin{equation}
\Gamma_{\rm D}(\vec{k})=-{\rm Im}\tilde{C}(\vec{k},M_{k,\omega})/2M_{k,\omega}  \,.
\end{equation}
This expression is valid if the $\omega$ and nontrivial $k$-dependence of 
$M_{k,\omega}$ is negligibly small, that is
\begin{equation}
\label{M0andk}
M_{k,\omega}^2=k^2+M^2_0\,,
\end{equation}
where $M_0$ is a constant.
In this study, $M_{k,\omega}$ is given by
\begin{equation}
M^2_{k,\omega}=k^2+m^2+e^2\frac{T^2}{6}-\tilde{f}_{A_0}(\omega,\vec{k})+\int \frac{d\omega'}{2\pi}{\rm P}\frac{1}{\omega-\omega'}i\tilde{C}_{\vec k}(\omega')\,.
\end{equation}
Both $\tilde{f}_{A_0}$ and the principal value integral are divergent.
In Appendix C, we show this divergence can be removed by renormalizing the scalar field strength.
In other words, we can cancel out this divergence by the counter term which is proportional to the kinetic term of the scalar field.
Although the $k$ and $\omega$-dependence of $M_{k,\omega}$ are non-trivial,
such corrections are proportional to $e^2$.
If we assume $M_{k,\omega}$ is given by Eq. (\ref{M0andk}), we find
\begin{align}
\label{dissipationratee2}
\Gamma_{\rm D}(\vec{k}) =&\frac{e^2}{4\pi}\frac{k}{\sqrt{M^2_0+k^2}}
\int^{p_f}_{p_i}dp\left(1+\frac{1}{e^{\beta p}-1}-\frac{1}{e^{\beta(\sqrt{M^2_0+k^2} -p)}-1} \right)   \notag\\
&\qquad \qquad \times\left[ -\frac{M^2_0}{k^2} + \frac{(M^2_0-m^2)\sqrt{M^2_0+k^2}}{k^2 p}  -\frac{(M^2_0-m^2)^2}{4k^2 p^2} \right]\\
p_i=&\frac{M^2_0-m^2}{2(\sqrt{M^2_0+k^2}+k)}\,,\quad p_f=\frac{M^2_0-m^2}{2(\sqrt{M^2_0+k^2}-k)}\notag
\end{align}
Though we have assumed $M_0>m$ in deriving the above expression, 
it is finite even in taking $M_0\rightarrow m$.
In this limit, we can obtain
\begin{equation}
\Gamma_{\rm D}(\vec{k}) =
\left\{
\begin{array}{ll}
\frac{e^2}{3\pi \beta}\frac{k^2}{m^2}\quad &k\ll m \\
\frac{e^2}{2\pi \beta} \quad &k\gg m \,.
\end{array}
\right.
\end{equation}
We show the dissipation rate as a function of $k$ in FIG. \ref{dissipationrate} for various values of
$\beta m$
\footnote{Note that we should not take the high-temperature limit ($\beta m\rightarrow0$) 
for the result in FIG. \ref{dissipationrate},
since in such a case the difference between $M_0$ and $m$ is not negligible.}.

Since both the dissipation and fluctuation come from the left diagram in FIG.\ref{diagram},
we can see the physical processes related to dissipation and fluctuation 
by cutting the diagram into two pieces \cite{Drewes:2013iaa}.
Considering the fact that a scalar boson 
cannot decay into a scalar boson of the same species and a massless gauge boson
due to energy and momentum conservation,
it may be doubtful that Eq.(\ref{dissipationratee2}) is the physical dissipation rate.
Though the dissipation rate shown in FIG.\ref{dissipationrate} is expressed as an integral over the loop momentum, 
only the $\vec{p}=\vec{0}$, or a soft photon loop, contributes to the resultant finite value.
From a mathematical point of view, it results from a cancellation between 
the divergent contribution from the bosonic distribution function 
and the vanishment of the phase space.
In order to see whether this finite result due to the aforementioned cancellation 
is physically relevant, we have considered the same problem
in a finite box having a spatial volume $V$ with a periodic boundary condition
where momentum is discretized and the zero-mode contribution is isolated.
It is found that the zero-mode contribution contains the thermal average of
the field value squared which evidently diverges since no particular field
value is energetically favored.
As a result, contribution to Eq.(\ref{dissipationratee2}) scales as $\Phi_\Lambda^2/V$,
where $\Phi_\Lambda$ is a cutoff of the zero-mode field amplitude.
Thus, the zero-mode contribution has an ambiguity arising from its
dependence on the order of taking the limit $\Phi_\Lambda \to \infty$ and $V \to \infty$.
Hence we may not trust the finite value obtained in Eq.(\ref{dissipationratee2}) which is based on the particular continuum calculation.
Indeed the Eq.(\ref{dissipationratee2}) itself would vanish if, we incorporate a plasmon mass to the gauge field using a dressed propagator,
or simply a mass term generated by a finite value of $\phi$.
In this case Eq.(\ref{noise kernel in kspace}) would also vanish, as it should.

\begin{figure}[!h]
 \begin{center}
  \includegraphics[width=60mm]{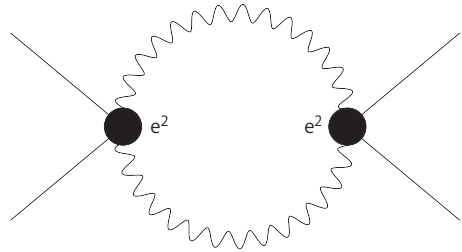}
 \end{center}
   \begin{minipage}[c]{140mm}
 \begin{spacing}{1}
 \caption{1PI diagram relevant to the multiplicative noise.
 This ${\mathcal O(e^4)}$ diagram leads to the nonzero dissipation rate for the coherently oscillating scalar field.
  \label{multi}}
  \end{spacing}
 \end{minipage}
\end{figure}

Thus the dissipation arises from diagrams higher order in $e$ as shown in FIG. \ref{multi}
related to the interaction $e^2A_{\mu}A^{\mu}\Phi^{\dagger}\Phi$.
In this case, the noise becomes the multiplicative noise, which appears in the equation of motion of $\phi$
in a form like $\xi \, \phi$.
The non-local memory term in the effective action is
\begin{align}
\Gamma\supset&-4ie^4\int d^4x_1 d^4x_2\left[\phi_{\rm cR}(x_1)\phi_{\rm \Delta R}(x_1)+\phi_{\rm cI}(x_1)\phi_{\rm \Delta I}(x_1)\right]
\left[|\phi_{\rm c}(x_2)|^2+\frac{1}{4}|\phi_{\rm \Delta}(x_2)|^2\right]\notag\\
&\quad\times \int \frac{d^3k_1}{(2\pi)^3}\frac{d^3k_2}{(2\pi)^3}e^{-i(\vec{k_1}+\vec{k_2})\cdot(\vec{x_1}-\vec{x_2})}
{\cal P}_{ij}(\vec{k_1}){\cal P}_{ij}(\vec{k_2})\notag\\
&\qquad\quad \left[g^>_{k_1}(t_1,t_2)g^>_{k_2}(t_1,t_2)-g^<_{k_1}(t_1,t_2)g^<_{k_2}(t_1,t_2) \right]\Theta(t_1-t_2)\,.
\end{align} 
The dissipation rate corresponding to multiplicative noise cases is also studied in Ref. \cite{Yokoyama:2004pf}.
Using the following quantity
\begin{align}
C_{\rm m}(x-x') \equiv &4ie^4\int \frac{d^3k_1}{(2\pi)^3}\frac{d^3k_2}{(2\pi)^3}e^{-i(\vec{k_1}+\vec{k_2})\cdot(\vec{x}-\vec{x'})}
{\cal P}_{ij}(\vec{k_1}){\cal P}_{ij}(\vec{k_2})\notag\\
&\left[g^>_{k_1}(t,t')g^>_{k_2}(t,t')-g^<_{k_1}(t,t')g^<_{k_2}(t,t')\right] \,,
\end{align}
we can evaluate the dissipation rate for the homogeneous field as 
\begin{equation}
\Gamma_{\rm D}=\frac{\tilde{C}_{\rm m}(\vec{k}=\vec{0},2M)}{2iM}|\phi(t)|^2=\frac{e^4|\phi(t)|^2}{4\pi M}(1+2N_{M})\,.
\end{equation}
Here $M$ is the angular frequency of the coherent oscillation 
and $|\phi(t)|^2$ is a mean square amplitude around the time $t$.
So even the coherent oscillation has nonzero dissipation at this order.

\section{Summary}
In this paper, we studied the role of gauge fields in the effective action for the scalar field
by considering the scalar QED theory.
As can be expected from previous studies,
the effective action we obtained contains an imaginary part.
We rewrote it by applying the Gaussian functional integral formula,
and interpreted the integral over variable $\xi$ as ensemble averaging.
The validity of this arrangement is confirmed by
the fluctuation-dissipation relation between 
the memory term and the introduced noise term.
Then we analyzed the spatial correlation of the noise,
and found the noise shows anti-correlation, which is different from the case of scalar and fermionic interactions.
The origin of this anti-correlation is due to the existence of derivative interactions 
between the scalar and gauge field.
We also considered the dissipation rate of the scalar field.
Though we obtained a finite dissipation rate, it comes from a soft photon in the loop.
It would vanish if we incorporate a finite mass which may be generated from higher order loops.
Furthermore
since the dissipation we have obtained comes from derivative interactions,
the dissipation rate for the coherent oscillation vanishes.
On the other hand higher order diagrams, consisting of a non-derivative interaction as depicted in FIG.\ref{multi},
gives a nonzero dissipation rate.

Considering that gauge coupling constants are generally larger than Yukawa coupling constants,
the absolute value of the noise correlation function for massless case (Eqs. (\ref{short}) and (\ref{long massless}))
can be larger than that of fermionic noise studied by Ref. \cite{Yamaguchi:1996dp}.
It would be interesting to study the phase transitions numerically
with our results included.
One of the other future works is to extend this study to non-Abelian gauge theories
in order to treat the realistic phenomena in the early Universe.

\section*{Acknowledgments}

We would like to thank Marco Drewes for helpful comments.
This work was supported by 
JSPS Research Fellowships for Young Scientists (Y.M.),
the Grant-in-Aid for Scientific Research on Innovative Areas 
No. 25103505 (T.S.), No. 24103006 (H.M.), No. 21111006 (J.Y.),
No. 25103504 (J.Y.) from 
The Ministry of Education, Culture, Sports, Science and Technology (MEXT),
and JSPS Grant-in-Aid for Scientific Research No. 23340058 (J.Y.).

\section*{Appendix A: short-range noise correlation }

From (\ref{alpha(r)})$\sim$(\ref{gammadoubleprime}),
we obtain the following asymptotic form as $r\rightarrow 0$.
\begin{align}
\alpha(r)&\rightarrow \frac{1}{2\pi^2r^2} \notag\\
\beta'(r)&\rightarrow -\frac{1}{2\pi^2 r}\,,\qquad
\beta''(r)\rightarrow  \frac{1}{2\pi^2 r^2}\notag\\
\gamma'(r)&\rightarrow -\frac{1}{\pi^2r^3}\,,\qquad
\gamma''(r)\rightarrow \frac{3}{\pi^2r^4}
\end{align}
To derive these results,
we have used the fact that modified Bessel functions $K_n(x)$ satisfy
\begin{equation}
\lim_{x\rightarrow 0}x^n K_n(x)=2^{-1+n}\Gamma(n)\,.
\end{equation}
Using the above expressions, the spatial noise correlation becomes
\begin{equation}
\label{Asymnoisecorr}
\langle \xi(\vec{x_1},t)\xi^{\dagger}(\vec{x_2},t)\rangle \,\simeq \,-\frac{3e^2}{2\pi^4r^6}\,,
\end{equation}
as $r$ approaches zero.

Since the value at $r=0$ corresponds to
$\langle |\xi(\vec{x},t)|^2 \rangle$, the correlation function given by Eq. (\ref{Asymnoisecorr})
being negative seems strange.
We speculate that the origin of this apparent contradiction lies in the evaluation of $\gamma_{ij}$. 
To see the essence, we now consider the case where the scalar field is massless.

The divergence comes from the zero-temperature part.
\begin{align}
\gamma'_{\rm zero}  &=\frac{1}{2\pi^2}\int_0^{\infty} dp  \left(\frac{p\cos(pr)}{r}-\frac{\sin(pr)}{r^2}\right) \\
\gamma''_{\rm zero} &=\frac{1}{2\pi^2}\int_0^{\infty} dp  \left(\frac{-p^2\sin(pr)}{r}-2\frac{p\cos(pr)}{r^2}+2\frac{\sin(pr)}{r^3}\right)
\end{align}
These integrals are UV divergent. 
We regulate them by introducing a cutoff factor $e^{-p/\Lambda}$, getting
\begin{align}
\gamma'_{\rm zero}  &=-\frac{1}{2\pi^2}\frac{2r\Lambda^4}{(1+r^2\Lambda^2)^2} \,,\\
\gamma''_{\rm zero} &=\frac{1}{2\pi^2}\frac{2\Lambda^4(3r^2\Lambda^2-1)}{(1+r^2\Lambda^2)^3}\,.
\end{align}
If we evaluate the noise correlation with these regulated integral,
the asymptotic form becomes
\begin{equation}
\langle \xi(\vec{x_1},t)\xi^{\dagger}(\vec{x_2},t)\rangle \,\simeq \,
-\frac{e^2}{2\pi^4}\frac{\Lambda^4(3r^2\Lambda^2-1)}{r^2(1+r^2\Lambda^2)^3}\,.
\end{equation}
When $r>0$, taking $\Lambda\rightarrow \infty$ gives the same result as (\ref{Asymnoisecorr}).
On the other hand, if we keep $\Lambda$ finite and take $r\rightarrow +0$, we see that the spatial noise correlation goes to $+\infty$.
At $r=0$, the dominant part is $\frac{e^2}{2\pi^4}\frac{\Lambda^4}{r^2(1+r^2\Lambda^2)^3}$.
If we multiply it by $r^5$ and integrate from $0$ to $\infty$, we obtain a finite value.
\begin{equation}
\int_0^{\infty}dr\, \frac{\Lambda^4}{r^2(1+r^2\Lambda^2)^3}\times r^5=\frac{1}{4}
\end{equation}
From this, we can write as follows.
\begin{equation}
\frac{e^2}{2\pi^4}\frac{\Lambda^4}{r^2(1+r^2\Lambda^2)^3}=\frac{e^2}{4\pi^4r^5}\delta(r)
\end{equation}

\section*{Appendix B: long-range noise correlation }

We briefly show the long-range ($r\gg\beta,\,\frac{1}{m}$) behavior.
In this limit, we obtain
\begin{align}
\alpha(r)&\rightarrow \frac{1}{2\pi\beta r}\notag\\
\beta'(r)&\rightarrow -\frac{1}{4\pi \beta}\notag\\
\beta''(r)&\rightarrow \frac{\beta}{12\pi r^3} \notag\\
\gamma'(r)&\rightarrow \left\{
\begin{array}{ll}
-\frac{1}{2\pi \beta r^2} &m=0 \\
-\frac{m}{2\pi r \beta}e^{-mr}& m\neq0
\end{array}
\right.\notag\\
\gamma''(r)&\rightarrow \left\{
\begin{array}{ll}
\frac{1}{\pi \beta r^3} &m=0 \\
\frac{m^2}{2\pi r \beta}e^{-mr}&m\neq0
\end{array}
\right.
\end{align} 

In the evaluation of $\gamma'$ and $\gamma''$, we used the asymptotic form for modified Bessel functions $K_n(x)$,
\begin{equation}
K_n(x)\rightarrow \sqrt{\frac{\pi}{2}}x^{-1/2}e^{-x}\quad {\rm as} \,\, x\rightarrow \infty .
\end{equation}

\section*{Appendix C: scalar field strength renormalization}

 We show the divergent part of
 \begin{equation}
 \label{appC1}
 -\tilde{f}_{A_0}(\omega,\vec{k})+\int \frac{d\omega'}{2\pi}{\rm P}\frac{1}{\omega-\omega'}i\tilde{C}_{\vec k}(\omega')
 \end{equation}
can be removed by renormalizing the field strength of the scalar field.

First, $\tilde{f}_{A_0}$ can be expressed as
\begin{equation}
\label{fA03d}
-\tilde{f}_{A_0}(\omega,\vec{k})=\frac{e^2}{2}\int \frac{d^3q}{(2\pi)^3}\frac{\omega^2+\omega_q^2}{\omega_q|\vec{k}-\vec{q}|^2}(1+2n_q)\,.
\end{equation}
This is an UV-divergent integral, whose divergence comes from zero temperature part.
Sticking to massless case which does not loss generality of the analysis in this section, we find
\begin{equation}
-\tilde{f}_{A_0}(\omega,\vec{k})\rightarrow \frac{e^2}{16\pi^2k}\int_0^{\infty} dq (\omega^2+q^2) \ln \frac{(k+q)^2}{(k-q)^2}\,.
\end{equation}
Now we use the dimensional regularization method. 
Changing the dimensions from $3$ to $3+\epsilon$ in Eq. (\ref{fA03d}) enables us to extract the divergence as follows.
\begin{equation}
-\tilde{f}_{A_0}(\omega,\vec{k}) = \frac{e^2}{12\pi^2}(3\omega^2+k^2)\frac{1}{\epsilon}+({\rm regular\,\, terms})
\end{equation}

Second, it is convenient to use another expression for the principal integral term,
\begin{align}
&\quad\int \frac{d\omega'}{2\pi}{\rm P}\frac{1}{\omega-\omega'}i\tilde{C}_{\vec k}(\omega')\notag\\
&=-\frac{e^2}{2}\int \frac{d^3p}{(2\pi)^3}{\cal P}_{ij}k_ik_j \frac{1}{p\,\omega_{k+p}} \notag\\
&\quad\times \Bigg[ (1+2N_p)\left( \frac{\rm P}{\omega+p+\omega_{k+p}} -\frac{\rm P}{\omega-p-\omega_{k+p}}-\frac{\rm P}{\omega+p-\omega_{k+p}}+\frac{\rm P}{\omega-p+\omega_{k+p}}\right)\notag\\
&\qquad +(1+2n_{k+p})\left( \frac{\rm P}{\omega+p+\omega_{k+p}} -\frac{\rm P}{\omega-p-\omega_{k+p}}-\frac{\rm P}{\omega-p+\omega_{k+p}}+\frac{\rm P}{\omega+p-\omega_{k+p}}\right)\Bigg]\,.
\end{align}
This is also UV divergent and we use the dimensional regularization method once more.
The above integral at large $p$ is simplified to
\begin{equation}
-\frac{e^2k^2}{3\pi}\int^{\infty}dp\,p^{-1+\epsilon}\rightarrow -\frac{e^2k^2}{3\pi^2}\frac{1}{\epsilon}\,.
\end{equation}

Finally we find Eq.(\ref{appC1}) diverges like $\frac{e^2}{4\pi^2}(\omega^2-k^2)\frac{1}{\epsilon}$.
This combination of $(\omega^2-k^2)$ ensures that we can remove this divergence by the renormalization of the scalar field strength.


\end{document}